\documentclass[5p,twocolumn,times,number]{elsarticle}

\usepackage{textcomp}
\usepackage{graphicx}
\usepackage{amsmath}
\usepackage[colorlinks,citecolor=blue,urlcolor=blue,linkcolor=blue]{hyperref}

\begin{document}

\begin{frontmatter}

\title{Characterization of a depleted monolithic pixel sensors in 150\,nm CMOS technology for the ATLAS Inner Tracker upgrade}

\author[CEA]{F.J.~Iguaz\corref{cor}}
\ead{iguaz@cea.fr}
\author[CEA]{F.~Balli}
\author[CPPM]{M.~Barbero}
\author[CPPM]{S.~Bhat}
\author[CPPM]{P.~Breugnon}
\author[Bonn]{I.~Caicedo}
\author[CPPM]{Z.~Chen}
\author[CEA]{Y.~Degerli}
\author[CPPM]{S.~Godiot}
\author[CEA]{F.~Guilloux}
\author[CEA]{C.~Guyot}
\author[Bonn]{T.~Hemperek\fnref{PIUB}}
\author[Bonn]{T.~Hirono\fnref{PIUB}}
\author[Bonn]{H.~Kr\"uger\fnref{PIUB}}
\author[CEA]{J.P.~Meyer}
\author[CEA]{A.~Ouraou}
\author[CPPM]{P.~Pangaud}
\author[Bonn]{P.~Rymaszewski\fnref{PIUB}}
\author[CEA]{P.~Schwemling}
\author[CEA]{M.~Vandenbroucke}
\author[Bonn]{T.~Wang}
\author[Bonn]{N.~Wermes\fnref{PIUB}}

\cortext[cor]{Corresponding author}

\address[CEA]{IRFU, CEA, Universit\'e Paris-Saclay, F-91191 Gif-sur-Yvette, France}
\address[Bonn]{University of Bonn, Nussallee 12, Bonn, Germany}
\address[CPPM]{Centre de physicque des particules de Marseille (CPPM), 163 Avenue de Luminy, Marseille, France}

\fntext[PIUB]{Also in Physikalisches Institut der Universit\"at Bonn, Nussallee 12, Bonn, Germany}

\begin{abstract}
This work presents a depleted monolithic active pixel sensor (DMAPS) prototype
manufactured in the LFoundry 150\,nm CMOS process.
DMAPS exploit high voltage and/or high resistivity inclusion of modern CMOS technologies
to achieve substantial depletion in the sensing volume.
The described device, named LF-Monopix, was designed as a proof of concept of a fully monolithic sensor
capable of operating in the environment of outer layers of the ATLAS Inner Tracker upgrade in 2025
for the High Luminosity Large Hadron Collider (HL-LHC).
This type of devices has a lower production cost and lower material budget compared to presently used hybrid designs.
In this work, the chip architecture will be described followed by the characterization of the different pre-amplifier
and discriminator flavors with an external injection signal and an iron source (5.9\,keV x-rays).
\end{abstract}

\end{frontmatter}

\section{Introduction}
\label{sec:Introduction}
\begin{figure}[htb!]
\centering
\includegraphics[width=0.90\linewidth]{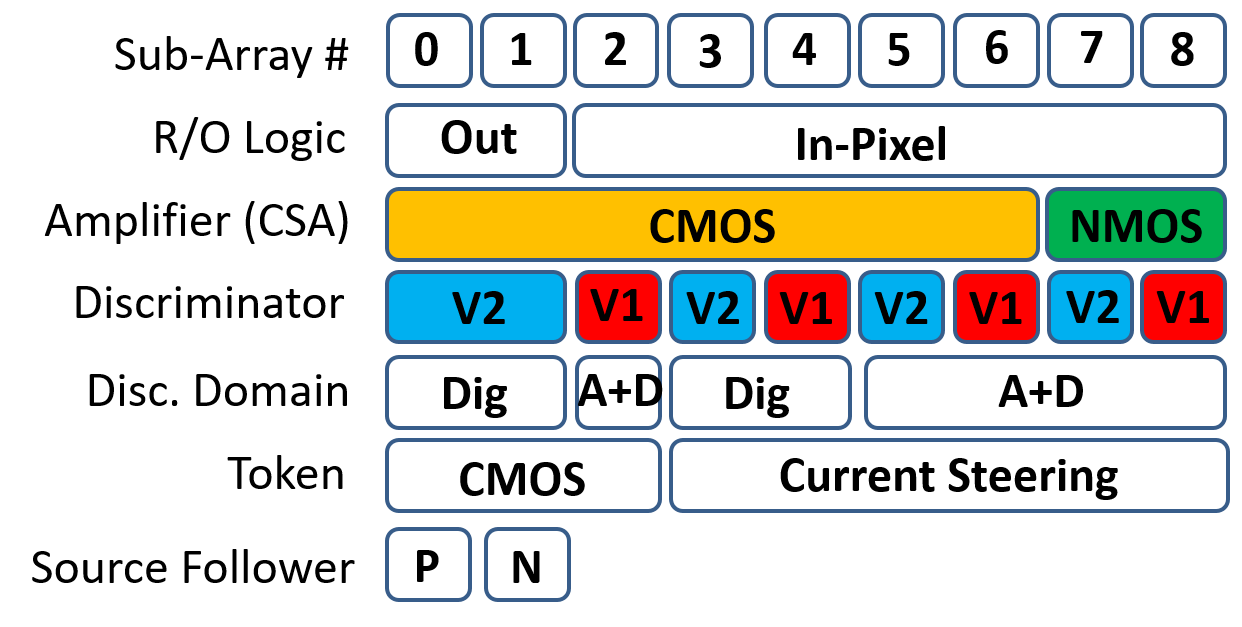}
\caption{Placement of pixel flavors in LF-Monopix chip}
\label{fig:ChipLayout}
\end{figure}

LF-Monopix chip~\cite{Wang:2016dgi,Wang:2017vlh} is the first fully monolithic prototype
of Depleted Monolithic Active Pixel Sensors (DMAPS)~\cite{Peric:2007zz}
implemented in the LFoundry 150\,nm CMOS technology aimed for high radiation environment.
Its design has kept significant features from its ancestor LF-CPIX~\cite{Degerli:2016sub}.
The chip size is $1 \times 1$\,cm$^2$, where most of the area is occupied by the pixel matrix composed of 36 columns and 129 rows.
The pixel size is 250\,\textmu m $\times$ 50\,\textmu m.
The pixel matrix is not uniform (see Fig.~\ref{fig:ChipLayout}) and includes nine pixel flavors (four columns per flavor).

The pixel types are composed of different combinations of pre-amplifier (NMOS or CMOS input),
discriminator (a two stage open-loop structure, called V1; and a self-biased differential amplifier with a CMOS output stage, called V2),
type of logic gate and placement of in-pixel readout logic.
This work describes the characterization of the chip performance of three samples (numbered 7, 11 and 12)
in terms of breakdown voltage, gain, input capacitance, threshold dispersion and noise.
LF-Monopix chip has also been tested under irradiation~\cite{Hirono:2018rqw}.


\section{Characterization of chip performance}
\label{sec:LabTests}
\begin{figure}[htb!]
\centering
\includegraphics[width=0.90\linewidth]{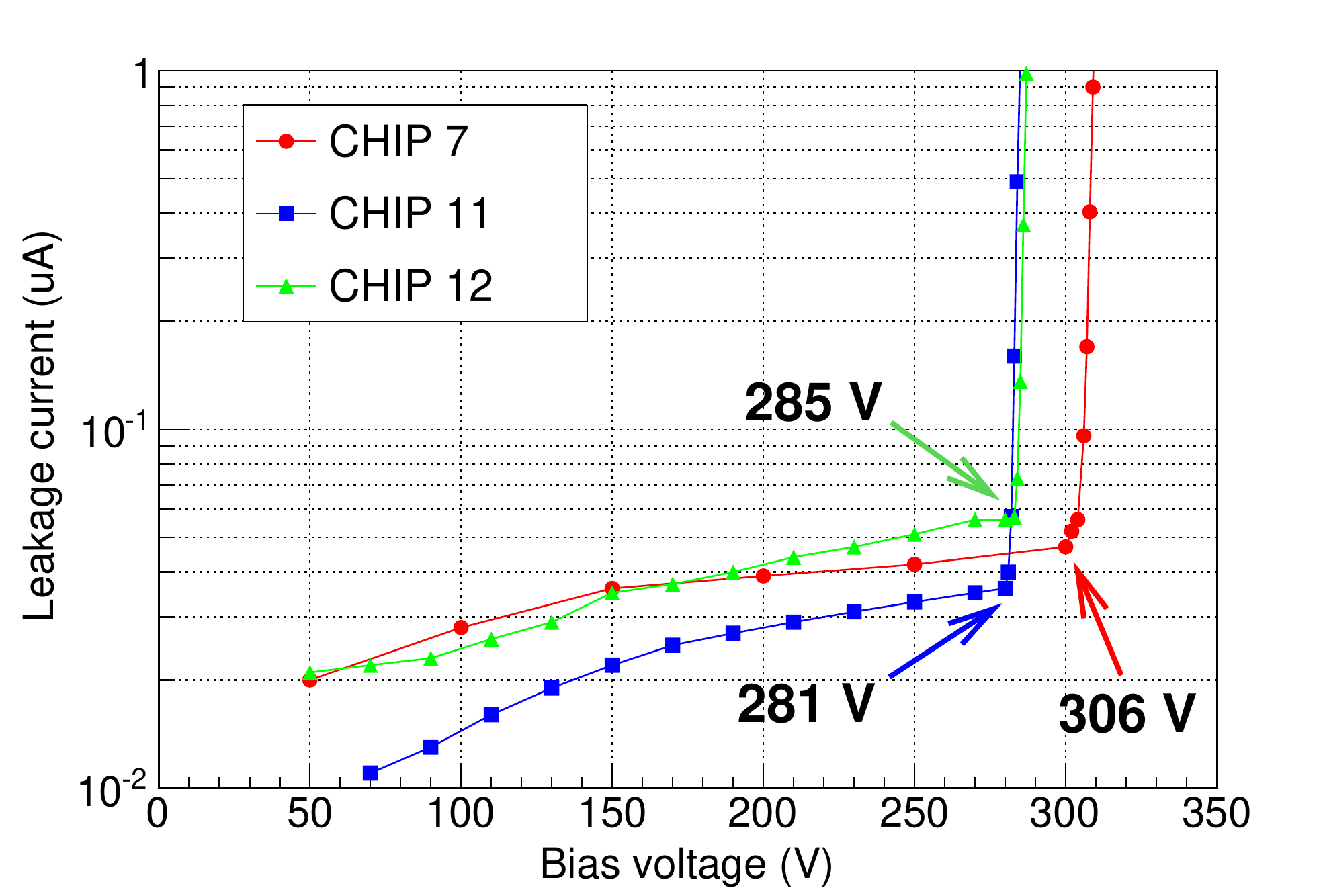}
\caption{Dependence of the leakage current on the bias voltage for three different samples of the LF-Monopix chip
at room temperature (20 degrees).}
\label{fig:BreakVoltage}
\end{figure}

The first measured parameter of LF-Monopix chip is the breakdown voltage,
which is the maximum voltage at which the detector can operate.
This value is derived from the dependence of the reverse bias voltage applied to the outer P-Well guard rings
with the measured leakage current, shown in Fig.~\ref{fig:BreakVoltage}.
Values for breakdown voltage between 280\, and 300\,V were obtained,
much higher than previous chip generation in the same technology~\cite{Hirono:2016pyl}.
This increase was possible by optimizing the guard ring structure based on TCAD simulations~\cite{Liu:2017dra}.

The gain of different pixels was calibrated using a $^{55}$Fe source (5.9\,keV x-rays)
and monitoring the response of individual pixels in the oscilloscope.
For each pixel, the histograms of the signal amplitude and the baseline were generated
and their mean values were measured.
The pixel gain was then calculated as the ratio of the difference between these mean values
and the mean number of ion-electron pairs generated in the substrate (1620\,electrons).
This procedure was repeated for a minimum of 8 pixels for each flavor.
As shown in Table~\ref{tab:ChipResults},
only a dependence on the type of preamplifier was observed.
Note that these gain values have to be considered as qualitative,
since the real gain of the analogue chip chain is not precisely known.
Nevertheless, the signal response of individual pixels can be used to estimate
the input capacitance, by calculating the injection signal that gives the same signal response.
The values obtained also show a dependence on the type of preamplifier.

\begin{table}[htb!]
\centering
\caption{The peformance of the two chip samples characterized in this work,
in terms of gain and input capacitance (C).}
\begin{tabular}{c|cc|cc}
\hline
\# & \multicolumn{2}{c|}{Gain ($\mu$V/e)} & \multicolumn{2}{c}{Capacitance (fF)}\\
  & CMOS & NMOS & CMOS & NMOS\\
\hline
7  & 15.9 $\pm$ 0.1 & 12.0 $\pm$ 0.1 & 2.40 $\pm$ 0.05 & 2.82 $\pm$ 0.05\\
12 & 15.5 $\pm$ 0.1 & 13.0 $\pm$ 0.1 & 2.25 $\pm$ 0.06 & 2.47 $\pm$ 0.08\\
\hline
\end{tabular}
\label{tab:ChipResults}
\end{table}


The threshold distribution and noise level are studied by scanning an external injection signal into the sensing node and
recording the probability of pixel firing with a fixed threshold setting (0.795\,V), obtaining the so-called ``S-curves''.
The threshold and noise parameters of each pixel are extracted by fitting an error function to the resulting S-curve.
The scan results are shown in Table~\ref{tab:SummaryFlavor}.
The threshold level is $\sim$2960 ($\sim$4300) electrons for the V1 (V2) discriminator
and its dispersion is $\sim$930 ($\sim$530) electrons.
Meanwhile, the noise level is $\sim$230 ($\sim$190) electrons for CMOS (NMOS) preamplifier.

\begin{table}[htb!]
\centering
\caption{The threshold, threshold dispersion and noise in electrons units before and after threshold tuning
for different types of discriminator (in the first two cases) and preamplifier (in the last one).}
\begin{tabular}{cc|cc}
\hline
Feature & Type & Untuned & Tuned\\
\hline
Threshold & V1 discri & 2960 $\pm$ 150 & -\\
(e-)      & V2 discri & 4290 $\pm$ 230 & -\\
\hline
Thr. Disp. & V1 discri & 533 $\pm$ 45 & 104 $\pm$ 14\\
(e-)       & V2 discri & 926 $\pm$ 89 & 153 $\pm$ 21\\
\hline
Noise      & CMOS      & 230 $\pm$ 17 & 238 $\pm$ 17\\
(e-)       & NMOS      & 193 $\pm$ 10 & 195 $\pm$ 13\\
\hline
\end{tabular}
\label{tab:SummaryFlavor}
\end{table}

\begin{figure}[htb!]
\centering
\includegraphics[width=0.90\linewidth]{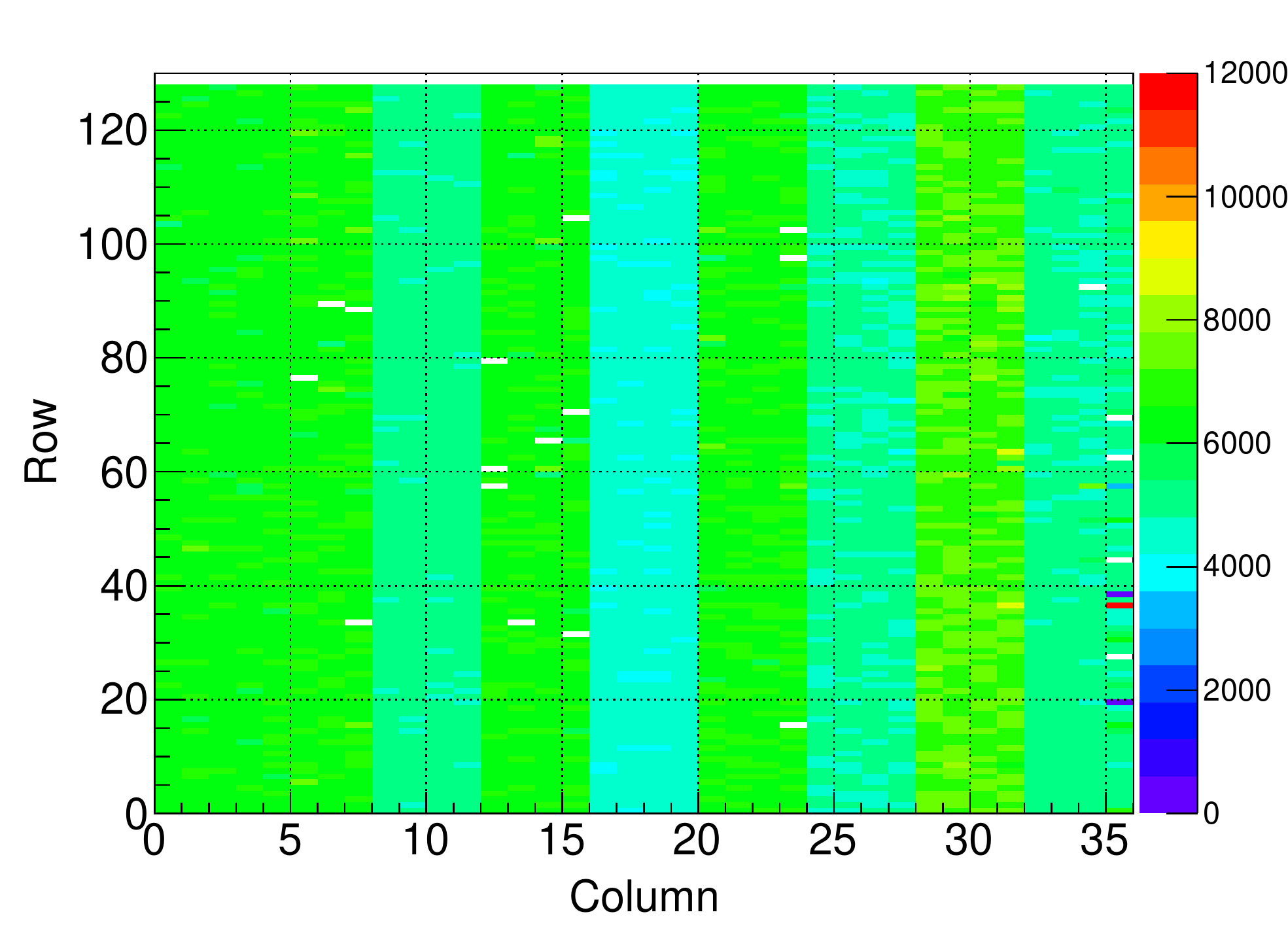}
\caption{Dependence of the threshold level in electrons units on the pixel position after the threshold tuning
for the sample 7 of the LF-Monopix chip.}
\label{fig:ThresholdTuning}
\end{figure}

The threshold of each pixel can be independently modified by a 4-bit in-pixel digital-to-analog converter (DAC) called TRIM.
Therefore, each pixel was tuned by scanning the TRIM values,
determining the corresponding threshold and fitting a linear relation between the TRIM values and the thresholds.
Then the optimal TRIM for each pixel is set such as to minimize the threshold dispersion.
The result of this procedure is shown for one chip in Fig.~\ref{fig:ThresholdTuning}.
The threshold dispersion is reduced down to $\sim$150 ($\sim$100) electrons for the V1 (V2) discriminator,
while noise levels remain the same.

\section{Summary}
\label{sec:Summary}
The performance of the different flavors of LF-Monopix chip has been characterized
with an external injection signal and an iron source.
The best results are obtained for the V1 discriminator
(threshold dispersion of $\sim$100 electrons) and the NMOS preamplifier (noise level of $\sim$190 electrons).

\section*{Acknowledgments}
This work has received funding from the European Union's Horizon 2020 Research and Innovation programme
under Grant Agreements no. 654168 (AIDA 2020) and 675587 (STREAM).
F.J.~Iguaz acknowledges the support from the Enhanced Eurotalents program (PCOFUND-GA-2013-600382).

\bibliographystyle{JHEP}
\bibliography{20180612_FJIguaz_LFMONOPIX}
\end{document}